\documentclass[sigconf]{acmart}
\AtBeginDocument{%
  }

\copyrightyear{2025}
\acmYear{2025}
\setcopyright{cc}
\setcctype{by}
\acmConference[EnvSys '25]{International Workshop on Environmental Sensing Systems for Smart Cities}{June 23--27, 2025}{Anaheim, CA, USA}
\acmBooktitle{International Workshop on Environmental Sensing Systems for Smart Cities (EnvSys '25), June 23--27, 2025, Anaheim, CA, USA}
\acmDOI{10.1145/3742460.3742984}
\acmISBN{979-8-4007-1986-8/2025/06}

\usepackage{enumitem}
\usepackage{float}
\begin{document}

\title{Localization using Angle-of-Arrival Triangulation}

\author{Amod K. Agrawal}
\authornote{This work extends research conducted independently by the author as part of the graduate program at the University of Illinois at Urbana-Champaign (UIUC).}
\email{amoagraw@amazon.com}
\orcid{0009-0001-9732-1024}
\affiliation{%
  \institution{Amazon Lab126}
  \city{Sunnyvale}
  \state{California}
  \country{USA}
}

\begin{abstract}
Indoor localization is a long-standing challenge in mobile computing, with significant implications for enabling location-aware and intelligent applications within smart environments such as homes, offices, and retail spaces. As AI assistants such as Amazon Alexa and Google Nest become increasingly pervasive, microphone-equipped devices are emerging as key components of everyday life and home automation. This paper introduces a passive, infrastructure-light system for localizing human speakers using speech signals captured by two or more spatially distributed smart devices. The proposed approach, \textit{GCC+}, extends the Generalized Cross-Correlation with Phase Transform (GCC-PHAT) method to estimate the Angle-of-Arrival (AoA) of audio signals at each device and applies robust triangulation techniques to infer the speaker’s two-dimensional position. To further improve temporal resolution and localization accuracy, feature-space expansion and subsample interpolation techniques are employed for precise Time Difference of Arrival (TDoA) estimation. The system operates without requiring hardware modifications, prior calibration, explicit user cooperation, or knowledge of the speaker’s signal content, thereby offering a highly practical solution for real-world deployment. Experimental evaluation in a real-world home environment yields a median AoA estimation error of $2.2^\circ$ and a median localization error of 1.25\,\text{m}, demonstrating the feasibility and effectiveness of audio-based localization for enabling context-aware, privacy-preserving ambient intelligence.
\end{abstract}



\maketitle
\captionsetup{skip=0pt}
\begin{figure}[h]
  \centering
  \includegraphics[width=\linewidth]{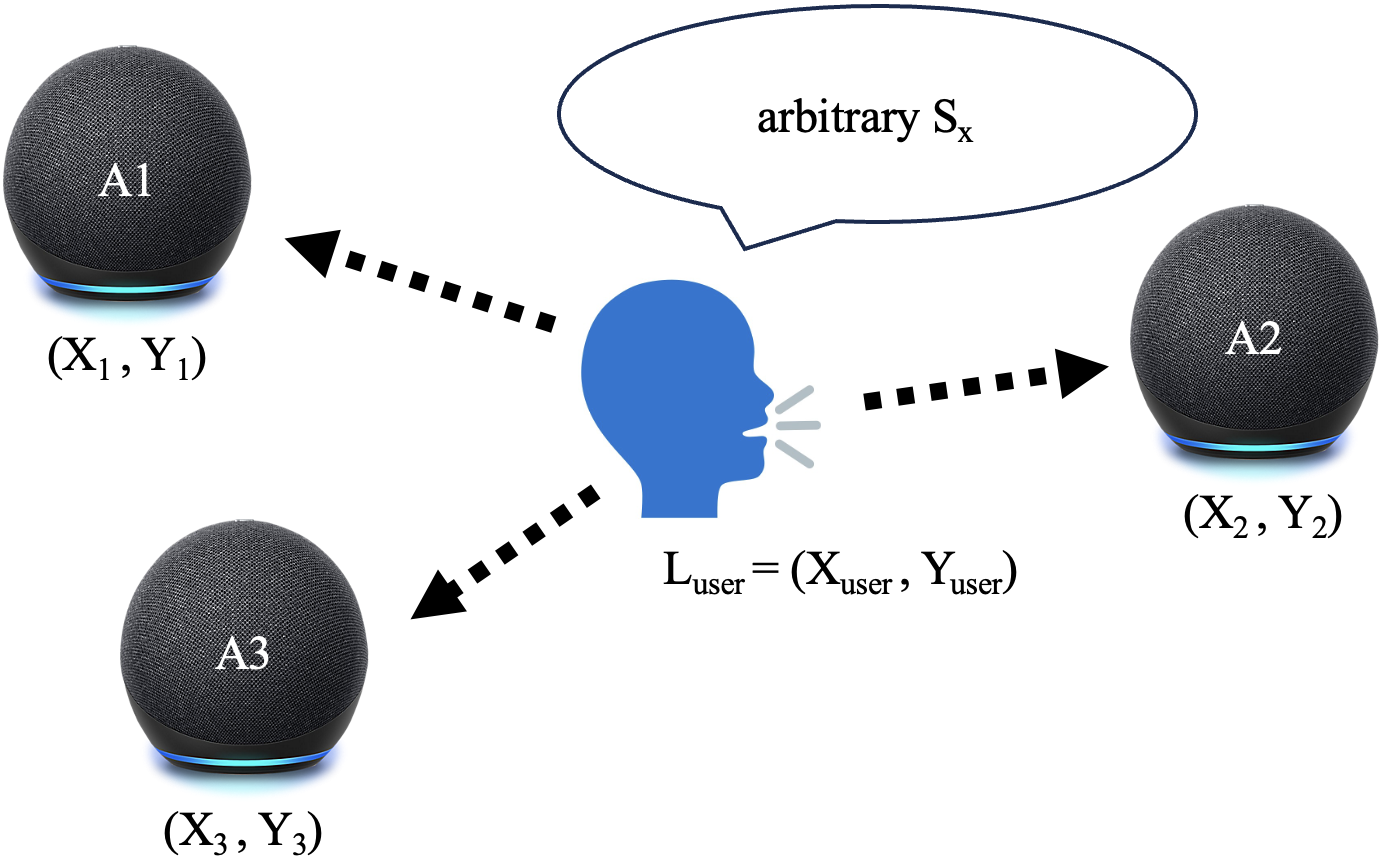}
  \caption{System setup showing three spatially distributed smart speakers equipped with microphone arrays}
  \label{fig:setup}
  \Description{The images showcases three Alexa devices named A1, A2, and A3 with marked coordinates. In the center of the image is a pictorial representation of a human being speaking an arbitrary sound signal. This sound signal is captured by three microphones arrays and is used to localize the user's unknown (x-y) coordinates.}
\end{figure}
\section{Introduction}
Indoor localization remains a fundamental research challenge in ubiquitous computing, context-aware systems, and smart environments. While GPS provides reliable global positioning, it is ineffective indoors due to signal attenuation and multipath propagation. Alternative approaches, such as vision-based systems, face limitations including occlusion, variable lighting conditions, and privacy concerns. RF-based approaches are susceptible to multipath effects and typically require users to carry active devices for ranging. In contrast, audio-based localization using commodity smart speakers presents a promising, low-cost, passive, and practical solution. 

As of 2024, more than 600 million smart home speakers have been sold globally, with 148 million units shipped in that year alone \cite{TechInsights2025}. With the widespread adoption of smart assistants such as Amazon Alexa and Google Nest, these devices are now widely distributed throughout the homes of users. Many households now deploy multiple devices across rooms, offering high spatial diversity and sensing capabilities. Leveraging their built-in microphones for passive localization enables the addition of contextual intelligence without requiring any additional infrastructure. Accurate indoor localization unlocks a range of applications, from home automation and energy optimization to safety and emergency response. However, significant challenges persist, including the need to robustly isolate human speech in the presence of noise, reverberation, and device-specific limitations.

We present a signal processing pipeline, \textit{GCC+}, for localizing a human speaker within an indoor environment using at least two smart speakers equipped with microphone arrays. By estimating the Angle-of-Arrival (AoA) of arbitrary speech signals at each device and triangulating these estimates, the system passively and accurately infers the speaker’s position. The approach is designed to be privacy-preserving, infrastructure-light, and deployable in real-world settings without requiring user cooperation, prior calibration, or hardware modifications. Real-world evaluations conducted in a standard home environment demonstrate improved AoA estimation accuracy, which directly contributes to enhanced localization performance over the baseline methods.
\vspace{-0.23em}
\section{Motivation}
To localize a user on a floor plan using AoA techniques, at least two fixed anchor nodes must be established as reference points within the environment. Modern smart home ecosystems increasingly provide tools to create digital floor plans for precise spatial configuration. For example, MapView \cite{AmazonMapView2025} offers an augmented reality (AR) experience on smartphones, enabling users to scan and reconstruct their home's layout, while LiDAR-equipped home robots can autonomously map floor plans during routine operation. If a spatial map is unavailable, device-to-device localization can establish a relative topology of devices within the environment. These methods allow smart devices to be positioned within a two-dimensional (x-y) coordinate space, either in absolute terms or relative to other devices. On-board inertial sensors (IMU) can determine the absolute orientation of the scanning device, translating the local reference frame to a consistent global frame. Together, these capabilities enable precise spatial localization of devices, opening new possibilities for accurate user positioning in indoor environments.

If smart assistants are capable of estimating user proximity within a room, they can dynamically tailor responses and adapt functionality based on the user’s spatial context. For example, when a command is issued from the kitchen, the assistant can prioritize contextually relevant actions such as controlling kitchen appliances, adjusting localized lighting, or providing recipe guidance, while deprioritizing devices in other rooms. Similarly, detection of user presence in the bedroom during nighttime hours could trigger behavior such as reducing notification volume, initiating a sleep routine, or dimming lights. These spatially aware behaviors improve the contextual relevance and responsiveness of smart assistants, fostering more natural and seamless user interactions.

Acoustic localization has been studied using a range of techniques, including time difference of arrival (TDoA), frequency-domain beamforming, and machine learning-based methods. Prior AoA-based systems, such as those in \cite{survey_ssl, localization_ssl_sr, grumiaux2022survey, van2019improving, echoloc}, often rely on specialized microphone arrays or high-end hardware to achieve fine-grained localization. Other approaches use neural networks \cite{vera2018end, Wang2017, Comiter2018} on spectrograms but these methods typically demand large, labeled datasets and struggle to generalize across different rooms and hardware configurations. Concerns about privacy leakage in collaborative sensing are not unique to ambient sensing, similar risks have been demonstrated in other domains \cite{Garcia2018}. In contrast, our system uses unmodified commodity devices and classical signal processing to achieve robust, privacy-preserving localization without calibration, training, or additional infrastructure.

\section{Problem Statement}
This paper addresses the problem of estimating a user's two-dimensional (2D) location within a room using acoustic signals captured by these spatially distributed devices. Specifically, the goal is to triangulate the 2D position of a human speaker using arbitrary speech signals received by the microphone array enabled devices, each with known coordinates. Figure~\ref{fig:setup} visualizes the system with a singular active speaker and an arbitrary acoustic source signal. In this work, the following assumptions are made:
\begin{enumerate}[leftmargin=*]
\item The speaker and all microphone arrays lie on a common two-dimensional (2D) plane.
\item The speech signal emitted by the user is unknown (i.e., non-cooperative source).
\item A single far-field source of signal (i.e., a single active speaker in a single occupant home).
\end{enumerate}
Given that existing infrastructure technologies cannot capture elevation, this study focuses on 2D localization, assuming devices at table height and relying solely on azimuthal angles for triangulation. Challenges such as multipath propagation and reverberation, which are inherent to indoor acoustic localization, are mitigated through signal processing techniques integrated into the localization pipeline.

Our primary contributions are: (1) GCC+, an enhanced acoustic pipeline that improves AoA estimation through frequency-based filtering, feature-space expansion, and subsample interpolation; (2) a design that requires no calibration or time synchronization across arrays, each array shares an independent clock; and (3) a portable method that is robust to variations in microphone specifications, frequency response, gain, and sensitivity; (4) source localization using either absolute or relative array placement, leveraging pre-existing spatial mapping or device-to-device ranging methods.
\section{System Configuration}
In this study, three off-the-shelf Amazon Echo devices, each featuring a circular array of six microphones, were employed to collect acoustic data. The devices were placed at known positions on a two-dimensional floor plan, and their coordinates were manually recorded to serve as anchor points for localization. Audio signals were sampled at $44.1\,\mathrm{kHz}$, with a single microcontroller distributing a shared clock to all microphones on a device to ensure time synchronization and phase alignment across channels. To evaluate system robustness, experiments were conducted in multiple spatial configurations by repositioning devices throughout the environment. Speech was passively recorded across multiple randomized speaker positions and orientations for each configuration.
\begin{figure}[h]
  \centering
  \includegraphics[width=\linewidth]{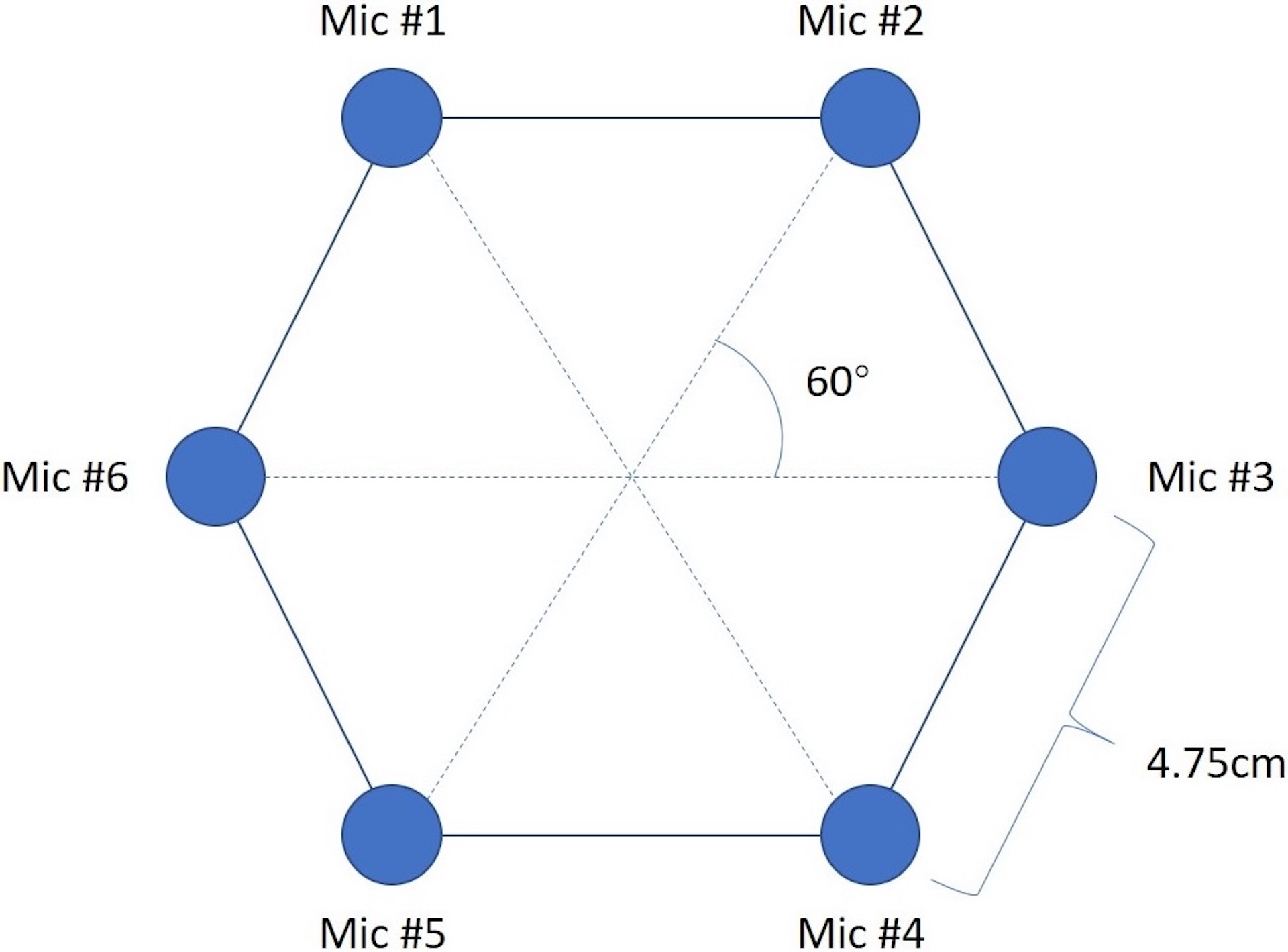}
  \caption{Hexagonal microphone array with $d = 4.75\,\mathrm{cm}$.}
  \label{fig:array}
  \Description{Image showcases a hexagonal microphone array with $d = 4.75\,\mathrm{cm}$.}
\end{figure}
Each microphone array consists of six elements arranged in a regular hexagonal formation with $60^\circ$ angular separation between adjacent microphones. The side length of the hexagon is $d = 4.75\,\mathrm{cm}$. \break Figure~\ref{fig:array} visualizes the microphone array. Given the center of the array at coordinates $(x,\,y)$ and its orientation angle $\theta$, the position of each microphone element is computed using the following equations:
{\setlength{\abovedisplayskip}{2pt}
 \setlength{\belowdisplayskip}{5pt}
\begin{align*}
x_{\text{mic}} &= x + d \cdot \cos(\theta) \\
y_{\text{mic}} &= y + d \cdot \sin(\theta)
\end{align*}
}
All experiments were carried out in a typical residential living room containing soft furnishings that introduced moderate reverberation. The walls were untreated, and the devices were placed on elevated flat surfaces at approximately uniform height, consistent with typical smart speaker usage in home environments. All signal processing and visualization were performed in MATLAB. Custom implementations of the Inverse Real Fast Fourier Transform (IRFFT) and Generalized Cross-Correlation with Phase Transform (GCC-PHAT) were developed to support algorithm exploration and were compared against MATLAB's built-in GCC-PHAT implementation for performance evaluation. Each array independently estimated the local angle-of-arrival (AoA) before transmitting the results to a central server for triangulation. Custom implementations of the wideband MUSIC algorithm and Maximum Likelihood Estimators for triangulation were used.

\section{Angle-of-Arrival (AoA) Estimation}
Although Angle-of-Arrival (AoA) estimation is a well-studied problem, accurately determining the AoA of an unknown speech signal using a compact microphone array remains a significant challenge due to the following factors.
\begin{enumerate}[leftmargin=*]
\item{\textbf{Multipath propagation and environmental noise:} Speech signals often arrive at the microphones along multiple indirect paths due to reflections and echoes, which, when combined with background noise, severely degrade the quality of the received signal.}
\item{\textbf{Limited array aperture:} The physical size of the microphone array is constrained, with a radius of only 4.75\,\text{cm}. This small aperture restricts the spatial resolution and impairs the array’s ability to distinguish between closely spaced sources.}
\item{\textbf{Insufficient spatial resolution at given sampling rate:} Accurate AoA estimation requires sub-millimeter resolution. At a sampling rate of 44.1\,\text{kHz} and assuming the speed of sound as 343 \,\text{m/s}, the spatial resolution achievable is approximately 7.77\,\text{mm}—several times coarser than required. As a result, conventional methods such as delay-and-sum perform poorly under these conditions.} 
\end{enumerate}

\subsection{Wideband MUSIC for AoA Estimation}

The Multiple Signal Classification algorithm (MUSIC) ~\cite{music} is a high-resolution, subspace-based technique for estimating the Angle-of-Arrival (AoA) of incident signals. It operates by decomposing the spatial covariance matrix of the received microphone array data into signal and noise subspaces using eigenvalue decomposition. AoA estimates are obtained by scanning over candidate directions and identifying those that are orthogonal to the noise subspace. Peaks in the resulting spatial spectrum $P_{\text{MUSIC}}(\theta)$ correspond to the estimated directions of arrival. Although originally formulated for narrowband signals, MUSIC can be extended to wideband sources through incoherent processing techniques such as Incoherent MUSIC (IMUSIC). In this method, the wideband signal is segmented into $K$ discrete frequency bins $\{f_1, f_2, \dots, f_K\}$, and the narrowband MUSIC algorithm is applied independently at each frequency. The spatial spectrum for each bin $k$ is computed in eq. \ref{eq:spatial_spectrum_music} as:

\begin{equation}
P_{\text{MUSIC}}^{(k)}(\theta) = \frac{1}{\mathbf{a}^H(f_k, \theta) \, \mathbf{E}_n^{(k)} \mathbf{E}_n^{(k)H} \, \mathbf{a}(f_k, \theta)}
\label{eq:spatial_spectrum_music}
\end{equation}

\noindent where $\mathbf{a}(f_k, \theta)$ is the array steering vector at frequency $f_k$ and direction $\theta$, and $\mathbf{E}_n^{(k)}$ denotes the noise subspace obtained from the eigenvalue decomposition of the spatial covariance matrix $\mathbf{R}_x^{(k)}$.

The final wideband AoA estimate is obtained by non-coherently averaging the spatial spectra across all selected frequency bins in eq. \ref{eq:avg_spatial_spectra}:
\begin{equation}
P_{\text{IMUSIC}}(\theta) = \frac{1}{K} \sum_{k=1}^{K} P_{\text{MUSIC}}^{(k)}(\theta)
\label{eq:avg_spatial_spectra}
\end{equation}

This approach captures AoA information across a broad frequency range without requiring subspace alignment across frequencies, thereby simplifying implementation. While MUSIC and its wideband extensions offer superior angular resolution, they are sensitive to environmental noise, microphone array imperfections, and multipath effects. These limitations can degrade its performance in dynamic and reverberant acoustic environments such as those found in homes or industrial settings.

\subsection{GCC-PHAT for TDoA and AoA Estimation}

The Generalized Cross-Correlation with Phase Transform (GCC-PHAT) \cite{Knapp1976} is used to estimate the time delay \( \tau \) between two microphone signals, \( x_1(t) \) and \( x_2(t) \). The cross-power spectrum is computed in the frequency domain in eq. \ref{eq:gcc_1} as:
\begin{equation}
    G_{12}(f) = X_1(f) X_2^*(f)
    \label{eq:gcc_1}
\end{equation}

where \( X_1(f) \) and \( X_2(f) \) are the Fourier transforms of the respective signals, and \( ^* \) denotes the complex conjugate. The PHAT weighting normalizes the magnitude in eq. \ref{eq:gcc_2}, retaining only the phase information:
\begin{equation}
    \Phi_{12}(f) = \frac{G_{12}(f)}{|G_{12}(f)|}
    \label{eq:gcc_2}
\end{equation}
The inverse Fourier transform of \( \Phi_{12}(f) \) gives the cross-correlation function in eq. \ref{eq:gcc_3}:
\begin{equation}
    R_{12}(\tau) = \int_{-\infty}^{\infty} \Phi_{12}(f) e^{j2\pi f \tau} df
    \label{eq:gcc_3}
\end{equation}
The estimated time delay \( \hat{\tau} \) corresponds to the time lag in eq. \ref{eq:gcc_4} that maximizes \( R_{12}(\tau) \):
\begin{equation}
    \hat{\tau} = \arg \max_{\tau} R_{12}(\tau)
    \label{eq:gcc_4}
\end{equation}
Assuming far-field propagation and known microphone geometry, the AoA \( \theta \) can be derived from \( \hat{\tau} \) using eq. \ref{eq:gcc_5}:
\begin{equation}
    \theta = \cos^{-1} \left( \frac{c \hat{\tau}}{d} \right)
    \label{eq:gcc_5}
\end{equation}
where \( c \) is the speed of sound and \( d \) is the distance between the two microphones.

In the GCC framework, the peak of the correlation spectrum corresponds to the line-of-sight (LOS) path delay, while smaller peaks may represent reflections due to multipath effects. By suppressing the dominance of low-frequency components, the phase whitening process sharpens the cross-correlation peak, resulting in more precise delay estimation. Compared to MUSIC, GCC-PHAT provides improved temporal resolution, particularly for wideband signals in arrays with limited spatial aperture.

\section{Algorithm Implementation Details}
The most error-prone component of our algorithm is the estimation of the Time Difference of Arrival (TDoA) between pairs of microphones. TDoA represents the time lag between audio signals captured by spatially separated sensors and is critical for accurate localization. The conventional approach for TDoA estimation involves time-domain cross-correlation. However, this method often yields poor results due to its sensitivity to noise and multipath propagation, and suffers from limited time resolution. To address these limitations, we employ the Generalized Cross-Correlation with Phase Transform (GCC-PHAT) \cite{Knapp1976}. Unlike standard cross-correlation, GCC-PHAT operates in the frequency domain, using the Fourier transforms of the input signals to construct the cross-power spectrum. The algorithm applies a phase transform that "whitens" the signal by normalizing its spectral power, thus emphasizing phase alignment over amplitude. This process enhances robustness to noise and reverberation, factors frequently encountered in real-world acoustic environments.

\begin{figure}[t]
  \centering
  \includegraphics[width=\columnwidth]{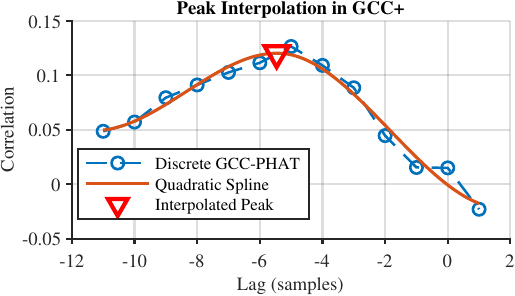}
  \caption{\textit{GCC+} leverages interpolation and quadratic splining to enable identification of peak at subsample level.}
  \Description{The plot illustrates how GCC+ enhances the original discrete GCC-PHAT correlation by applying quadratic spline interpolation. This technique fits a smooth curve through the discrete points, enabling precise estimation of the peak location at a subsample resolution. The interpolated peak provides a more accurate estimate of the time delay compared to using the discrete peak alone.}
  \label{fig:peak_interpolation}
\end{figure}

The proposed algorithm for Angle-of-Arrival (AoA) estimation and source localization proceeds through the following steps:

\begin{enumerate}[leftmargin=*]

\item \textbf{Signal Preprocessing:} Raw microphone signals are band-pass filtered and normalized to suppress irrelevant frequencies and enhance speech content. Since the captured audio signal is a wideband signal and consists of frequencies as high as 22.05\,\text{kHz}. However, most of the human speech lies in the spectrum of 300\,\text{Hz} to 3.5\,\text{kHz}, which is an important factor that can be used to discard frequencies that can pollute the AoA correlation.

\item \textbf{Delay Estimation using GCC-PHAT:} Compute the Time Difference of Arrival (TDoA) between all pairs of microphones using the Generalized Cross-Correlation with Phase Transform (GCC-PHAT), which enhances robustness to reverberation and wideband noise. Instead of selecting a single reference microphone and computing delays relative to it---yielding only five measurements for a six-microphone array, we compute TDoAs for all possible microphone pairs, resulting in \( \binom{6}{2} = 15 \) unique pairs. Furthermore, by considering symmetric pairs, directional delays, and multiple time windows, the effective dimensionality of the delay vector increases significantly. In our implementation, this produces a 36-dimensional delay vector, expanding the representation space from \( \mathbb{R}^5 \) to \( \mathbb{R}^{36} \). This higher-dimensional encoding captures more spatial structure and increases resilience to noise and outliers in AoA estimation.

\item \textbf{Subsample Interpolation:} Apply closed-form quadratic interpolation near the GCC-PHAT peak to obtain subsample TDoA precision, which improves angular resolution. Figure~\ref{fig:peak_interpolation} illustrates peak correction using subsample interpolation.

\item \textbf{AoA Estimation:} Use all pairwise TDoAs to compute a high-dimensional displacement vector for each microphone array. Perform a grid search over azimuth angles (0° to 360°) to identify the angle that best matches the observed vector. Visualize the AoA spectrum as a heatmap.

\item \textbf{Triangulation Setup:} Use the AoA estimates from at least two spatially separated microphone arrays with known anchor positions to construct direction lines. A third array may optionally be used to improve localization robustness.

\item \textbf{Robust Localization:} When only two microphone arrays are available, we localize the speaker using Maximum Likelihood Estimation (MLE), which in this case reduces to a least-squares formulation based on the intersection of two directional rays. The availability of more than two arrays yields an overdetermined system, which provides an opportunity to improve localization using more sophisticated solvers.
\end{enumerate}

\subsection{Improving AoA Accuracy with Interpolation}
To improve localization accuracy, we refine the Time Difference of Arrival (TDoA) estimates using subsample interpolation around the correlation peak. The raw correlation output is first upsampled by zero-padding between frequency components, followed by a low-pass filter. This results in an interpolated cross-correlation function with increased temporal resolution.
\captionsetup{skip=0pt}
To achieve subsample precision in TDoA estimation, we apply quadratic spline interpolation over a 6-point window centered around the correlation peak. This technique fits a second-order polynomial to the local region of the correlation curve, allowing for smooth and accurate peak refinement. The quadratic function is modeled in eq. \ref{eq:interpolation} as:
\begin{equation}
    f(t) = at^2 + bt + c,\quad\delta t = -\frac{b}{2a}
    \label{eq:interpolation}
\end{equation}
\
where \( t \) represents the sample index relative to the center of the window, and \( f(t) \) denotes the correlation value at that point. The coefficients \( a, b, c \) are obtained via least-squares fitting over the six correlation values. The refined delay estimate \( \delta t \) is given by the vertex of the fitted parabola, which is then added to the original integer-valued peak index to obtain a high-resolution TDoA estimate.
\captionsetup{skip=0pt}
 \begin{figure}[t]
  \centering
  \includegraphics[width=\columnwidth]{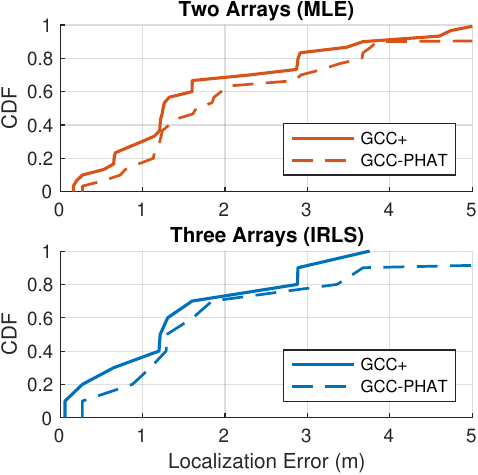}
  \caption{Performance comparison of \textit{GCC+} and the baseline GCC-PHAT.}
  \Description{The plots showcase performance comparison of two methods of localization: GCC+ and GCC-PHAT, by comparing their localization error's Cumulative Distribution Function (CDF). In both, two arrays and three arrays set-up, GCC+ outperforms GCC-PHAT.}
  \label{fig:localization_cdf}
\end{figure}
\subsection{Localization via Triangulation}
To estimate the speaker's location, we leverage AoA estimates \( \theta_i \) obtained from at least two microphone arrays with known anchor positions \( a_i \in \mathbb{R}^2 \). Each AoA defines a line in the two-dimensional plane and its parametric form is given by eq. \ref{eq:line}:
\begin{equation}
    \ell_i(t) = a_i + t \cdot n_i, \quad -\infty < t < \infty
    \label{eq:line}
\end{equation}
where \( n_i = [\cos\theta_i, \sin\theta_i]^T \) is the unit direction vector for array \( i \), and \( a_i \) denotes the known position of the array.
Ideally, the intersection of these lines would indicate the true source location. However, due to noise in AoA estimation and hardware imperfections, these lines often fail to intersect at a single point. To resolve this, we employ three solvers to estimate the location given observed AoAs.

\subsubsection{Maximum Likelihood Estimation}
MLE \cite{mle} assumes that AoA errors are independently and Gaussian distributed. It computes the source location by minimizing the sum of squared perpendicular distances from a candidate point to each AoA line. This approach is efficient and admits a closed-form solution, making it effective in scenarios with low measurement noise.

\subsubsection{Random Sample Consensus}
RANSAC \cite{ransac} offers robustness against outliers and performs effectively in noisy environments. The algorithm iteratively samples random pairs of AoA lines to generate candidate locations and evaluates each based on the number of consistent measurements. The candidate with the highest consensus is selected, enabling reliable localization even in the presence of inaccurate data.

\subsubsection{Iteratively Reweighted Least Squares}
IRLS \cite{IRLS} combines the strengths of MLE and RANSAC by assigning reliability-based weights to each AoA measurement. It iteratively updates these weights, attenuating the influence of outliers without discarding any data. This yields a robust estimate that effectively balances accuracy and computational efficiency.

We evaluate all three algorithms for accuracy and robustness. While MLE offers speed, RANSAC is particularly effective in the presence of outliers, and IRLS strikes a balance between the two. Although localization is feasible with only two arrays, incorporating a third improves accuracy and robustness. In our experiments, all methods demonstrate comparable performance under a three-array configuration; therefore, we report results for IRLS to minimize visual redundancy. When limited to two arrays, all methods converge to a least-squares (MLE) solution.

\section{Evaluation}
We evaluate the performance of our AoA estimation and end-to-end localization pipeline using 50 ground-truth samples collected in a natural home environment. Recording conditions include moderate background noise such as refrigerator hum, ambient music, road traffic, and fan noise. As the devices require wall-connected power, they are typically placed near walls on wooden tables, reflecting real-world smart speaker deployments. The distance between the speaker and microphone arrays varies from 0.47~\text{m} to 5.2~\text{m}. The speaker utters a single word, such as ``Hello'' with an average signal length of 1.06~\text{s}. Ground-truth locations were measured manually using a laser rangefinder, and the corresponding ground-truth angles were computed using trigonometric calculations along with a digital compass. Localization error is quantified using the Euclidean (\( \ell_2 \)) norm between the predicted location and the ground-truth location.
\captionsetup{skip=0pt}
\begin{figure}[t]
  \centering
  \includegraphics[width=\columnwidth]{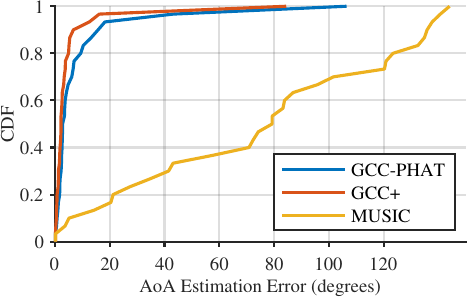}
  \caption{Performance comparison of \textit{GCC+}, GCC-PHAT, and wideband MUSIC}
  \Description{The CDF plot compares the AoA estimation error across three algorithms: GCC-PHAT, GCC+, and MUSIC. GCC+ (red) consistently outperforms GCC-PHAT (blue), offering improved accuracy by reducing the estimation error across most of the distribution. MUSIC (yellow), in contrast, exhibits significantly higher error, with a large portion of estimates deviating far from the true angle, indicating poor performance in this scenario. This demonstrates the effectiveness of \textit{GCC+} in achieving robust subsample AoA estimation.}
  \label{fig:aoa_cdf}
\end{figure}
The proposed system, \textit{GCC+}, achieves a median localization error of 1.25\,\text{m} and a mean error of 1.58\,\text{m}. Since localization performance is closely tied to the precision of AoA estimation, we also report angular estimation errors. The median AoA error is $2.2^\circ$, while the mean AoA error is $5.9^\circ$. The system operates online and in real time, with the full pipeline completing within the duration of a typical voice command (\textasciitilde1200\,\text{ms}), ensuring low-latency localization suitable for interactive smart home applications. Importantly, audio recordings can be discarded immediately after AoA estimation, thereby preserving user privacy. Table~\ref{tab:aoa_error} summarizes the AoA estimation error in degrees, and Figure~\ref{fig:aoa_cdf} presents the cumulative distribution function (CDF) of the AoA error.
\begin{table}
  \caption{AoA Estimation Error (Degrees)}
  \label{tab:aoa_error}
  \begin{tabular}{lccc}
    \toprule
    AoA Algorithm & Mean & Median & 90th \%\\
    \midrule
    MUSIC & 76.2 & 79.2 & 136.7\\
    GCC-PHAT & 9.4 & 3.1 & 17.1\\
    \textbf{GCC+} & \textbf{5.9} & \textbf{2.2} & \textbf{9.6}\\
  \bottomrule
\end{tabular}
\end{table}
\begin{table}
  \caption{(\( \ell_2 \)) norm Localization Error (Meters)}
  \label{tab:loc_error}
  \centering
  \begin{tabular}{l l c c c}
    \toprule
    Setup & Method & Mean & Median & 90th \% \\
    \midrule
    2 Arrays (MLE) & GCC-PHAT & 3.62 & 1.74 & 8.48 \\
                   & \textbf{GCC+}     & \textbf{1.83} & \textbf{1.25} & \textbf{4.13} \\
    3 Arrays (IRLS) & GCC-PHAT & 2.84 & 1.45 & 8.40 \\
                   & \textbf{GCC+}     & \textbf{1.58} & \textbf{1.25} & \textbf{3.32} \\
    \bottomrule
  \end{tabular}
\end{table}
Table~\ref{tab:loc_error} reports the localization error in meters, and Figure~\ref{fig:localization_cdf} illustrates the CDF of localization errors under two-array and three-array configurations.

\section{Discussion and Future Work}
\textit{GCC+} integrates a variation of GCC-PHAT with triangulation techniques to estimate a user's location based on arbitrary sound signals. However, this approach operates under several assumptions that may not hold in all real-world scenarios.
First, the algorithm assumes a single active speaker within the environment, limiting its effectiveness in multi-occupant households. Speaker profiling techniques that extract acoustic features and identify speaker-specific signatures \cite{voiceid} can distinguish overlapping speech signals, enabling independent angle-of-arrival (AoA) estimation for multiple users. Second, as microphone-equipped devices are increasingly deployed at varying heights in domestic environments, we plan to explore 3D localization by leveraging azimuth estimates from multiple planar arrays positioned at different elevations, enabling elevation inference via geometric triangulation without the need for fully non-coplanar array configurations. Lastly, the current approach assumes prior knowledge of anchor node locations through integration with existing smart home infrastructure. In practice, users may introduce errors when marking device locations. To address this, future work may explore device-to-device localization via acoustic ranging or complementary RF-based techniques such as BLE or Ultra-Wideband, enabling autonomous anchor localization, reducing user burden, and improving overall system reliability. Future work will expand testing across a broader range of indoor environments (e.g., office spaces, open-plan areas) and with a more diverse set of smart speakers to assess cross-device robustness. Although the system demonstrated resilience to environmental noise, a systematic evaluation of the failure point by artificially injecting controlled noise into the input signals would provide insights into the limits of noise tolerance.

\section{Conclusion}
This paper presents \textit{GCC+}, a practical and robust system for passive indoor localization using unmodified, commodity smart home devices. It integrates a refined Generalized Cross-Correlation with Phase Transform (GCC-PHAT) method for angle-of-arrival (AoA) estimation with a triangulation framework based on Maximum Likelihood Estimation (MLE). The system enables accurate, real-time localization in real-world environments without hardware modifications or calibration, and integrates seamlessly with existing smart home ecosystems. Notably, it reduces the 90th percentile AoA error from $17.1^\circ$ to $9.6^\circ$, and the corresponding localization error from 8.40\,\text{m} to 3.32\,\text{m}, demonstrating consistent performance under challenging acoustic conditions. \textit{GCC+} lays the groundwork for context-aware environments and future extensions such as multi-user tracking, 3D localization, and self-calibrating device networks.
\vspace{-1em}
\bibliographystyle{ACM-Reference-Format}
\bibliography{citations}
\end{document}